\documentclass[12pt,a4paper,dvips]{article}
\usepackage{cite,mcite}
\usepackage{graphicx}
\graphicspath{{/l3/paper/example/figs/}}
\newlength{\capindent}
\newlength{\capwidth}
\newlength{\figwidth}
\newcommand{\icaption}[2][!*!,!]{\hspace*{\capindent}%
  \begin{minipage}{\capwidth}
    \ifthenelse{\equal{#1}{!*!,!}}%
      {\caption{#2}}%
      {\caption[#1]{#2}}
  \end{minipage}}
\newcommand {\Be}{\begin{equation}}
\newcommand {\Ee}{\end{equation}}

\newcommand {\eqref}[1]{equation~(\ref{#1})}

\newcommand {\Figref}[1]{Figure~\ref{fig:#1}}

\renewcommand{\thefootnote}{\fnsymbol{footnote}}

\begin{document}

\begin{titlepage}
\begin{flushright} 
hep-ph/9907380 \\
July  14, 1999 
\end{flushright}

\vspace*{3.0cm}

\begin{center} {\Large \bf
       Global Analysis of Bhabha Scattering at LEP2\\
\vspace*{0.15cm}
       and Limits on Low Scale Gravity Models}

\vspace*{2.0cm}
  {\Large
  Dimitri Bourilkov\footnote{\tt e-mail: Dimitri.Bourilkov@cern.ch}
  }

\vspace*{1.0cm}
  Institute for Particle Physics (IPP), ETH Z\"urich, \\
  CH-8093 Z\"urich, Switzerland
\vspace*{4.3cm}
\end{center}

%
%
%
\begin{abstract}
A global analysis of the data on Bhabha scattering from the four
LEP experiments ALEPH, DELPHI, L3 and OPAL is performed to search
for effects of virtual graviton exchange in models with low scale
gravity. No statistically significant deviations from the Standard
Model expectations are observed and lower limits on the 
scale of models with large extra dimensions of
$\rm \Lambda_T = 1.077\ TeV$ for $\lambda = -1$ and
$\rm \Lambda_T = 1.412\ TeV$ for $\lambda = +1$
at 95~\% confidence level are derived.
\end{abstract}

\vspace*{1.0cm}

\end{titlepage}

\renewcommand{\thefootnote}{\arabic{footnote}}
\setcounter{footnote}{0}
%
%
\section*{Introduction}

The Standard Model (SM) has been tremendously successful when
confronting the theory with data coming from the highest energy
accelerators. Still, we believe that it is not complete, and
one of the first questions in all searches for new physics is
what is the relevant scale, where new phenomena will become
accessible to experiments. Recently, a radical proposal has been
put forward by Arkani-Hamed, Dimopoulos and Dvali~\cite{ADD}
for the solution of the hierarchy problem, which brings close
the electroweak scale $\rm m_{EW} \sim 1\; TeV$ and the
Planck scale $\rm M_{Pl} = \frac{1}{\sqrt{G_N}} \sim 10^{15}\; TeV$.
In this framework the effective 4 dimensional $\rm M_{Pl}$ is
connected to a new $\rm M_{Pl(4+n)}$ scale in a (4+n) dimensional
theory:
\Be
\rm M_{Pl}^2 \sim M_{Pl(4+n)}^{2+n} R^n
\Ee
where there are n extra compact spatial dimensions of radius
$\rm \sim R$.
Putting $\rm M_{Pl(4+n)} \sim m_{EW}$, 
for $\rm n = 1$ we get $\rm R \sim 10^{13}\; cm$, which is
excluded experimentally, but already
for $\rm n = 2$ the result is $\rm R \sim 0.1-1\; mm$, which
is below the current experimental limits from gravitational
experiments.

In this work we will adopt the notation from~\cite{Giudice} and
call the gravitational mass scale $\rm M_D$. This scale is relevant
in direct searches for graviton production. In the case of virtual
graviton exchange, which is the subject of our analysis, the
corresponding scale is the ultraviolet cutoff energy, denoted
$\rm \Lambda_T$ in~\cite{Giudice} and $\rm M_s$ in~\cite{Hewett,Rizzo},
with the relation $\rm M_s=(2/\pi)^{1/4}\Lambda_T$.
The two scales are expected to be close to each other.

This paper is organized as follows. In section 2, we describe the
effects of virtual graviton exchange in Bhabha scattering, which
turns out to be the single most sensitive channel at LEP2 energies.
In the following section the experimental data used in this analysis
is presented. In section 4 the results of the global fit are shown
and limits on low scale gravity models are derived.

%
%
\section*{Virtual Graviton Exchange in Bhabha Scattering}

The differential cross section for fermion-pair production
in $\rm e^+e^-$ collisions can be decomposed in the usual
way as:
\Be
\rm \frac{d \sigma}{d \Omega} = SM(s,t)+\varepsilon\cdot C_{Int}(s,t)+\varepsilon^2\cdot C_{Graviton}(s,t)
\Ee
where $\rm SM(s,t)$ is the Standard Model contribution,
$\rm C_{Graviton}(s,t)$ comes from the virtual graviton exchange and
$\rm C_{Int}(s,t)$ is the interference between the SM and the low
scale gravity terms. The exact form of these functions is given
in~\cite{Giudice} for all final states, in~\cite{Hewett} for 
final states other than electrons and in~\cite{Rizzo} for 
Bhabha scattering. Here we will use the results from the calculations
by Rizzo. The independent theoretical calculations in~\cite{Giudice} give
numerically values which are very close and produce the same final
results.

In the formula above
\Be
\rm \varepsilon = \frac{\lambda}{M_s^4}.
\Ee
The coefficient $\rm \lambda$ is of $\rm \mathcal{O}(1)$ and can not be
calculated explicitly without knowledge of the full quantum gravity
theory. In the following analysis we will assume that
$\rm \lambda = \pm 1$ in order to study both the cases of positive
and negative interference.

It should be noted that the exchange of a spin 2 particle leads
to terms $\rm \sim \cos^3\theta$ and $\rm \sim \cos^4\theta$, which
makes the differential cross sections a unique signature for this
type of physics. For fermions other than electrons the integrated
interference term for scattering angles from zero to $\rm \pi$
is exactly zero, and the graviton
exchange is suppressed by $\rm 1/M_s^8$. On the contrary, the
interference between the graviton exchange and t-channel SM exchanges
for Bhabha scattering is giving sizeable contributions, making the
$\rm e^+e^-$ final state the most sensitive search field. This
combines favourably with the larger cross section and the much
higher statistical precision of this measurement. In the area of
the forward peak the theory uncertainty in the SM predictions is
the limiting factor in our study.

Initial-state radiation (ISR)
changes the effective centre-of-mass energy in a large fraction of the
observed events.
We take these effects into account by computing the first order
exponentiated cross sections and asymmetries following~\cite{Kleiss}.
Other QED corrections give smaller effects and are neglected.

%
%
\section*{Experimental Data}

All LEP collaborations have submitted for publication or send to
recent conferences their measurements of fermion-pair production
at 183 and 189 GeV centre-of-mass energies. As explained in the
previous section, the golden channel to search for virtual
graviton exchange effects is Bhabha scattering and in the
following we will concentrate on the data for these two
highest energy points, where large data samples have been
accumulated during the very successful LEP runs in 1997 and 1998.

The DELPHI~\cite{de189} and L3~\cite{fpp189} collaborations have
presented preliminary results for total cross sections and
forward-backward asymmetries in the angular range
$\rm  44^{\circ}<\theta<136^{\circ}$, where $\rm \theta$ is
the angle between the incoming and the outgoing electrons.
The ALEPH~\cite{al183,al189} and OPAL~\cite{op183,op189}
collaborations have
presented results for the differential cross sections in
the angular range $\rm |\cos\theta|<0.9$. The scattering
angle is defined by OPAL to be in the laboratory frame, and
by ALEPH to be in the outgoing $\rm e^+e^-$ rest frame.
The experiments use different strategies to isolate the
high energy sample, where the energy of the propagator is
close to the full available centre-of-mass energy. This
sample is the main search field for new physics.
DELPHI and OPAL apply an acolinearity cut of $\rm 20^{\circ}$
and $\rm 10^{\circ}$ respectively.
ALEPH defines the effective energy, $\rm s'$, as the square
mass of the outgoing fermion pair. It is determined from the
angles of the outgoing fermions. L3 defines $\rm s'$ as the
mass squared of the $\rm \gamma^{*}/Z$ propagator. It is
determined from the invariant mass squared of the final state
$\rm e^+e^-$ pair, computed from the energies measured in
the electromagnetic calorimeter. Close-by final state radiation
photons are absorbed in the same energy cluster, and consequently
included in the invariant mass.

For details of the selection procedures, the statistical and
systematic errors we refer the reader to the publications of
the four LEP experiments.

%
%
\section*{Results and discussion}

The Standard Model predictions for Bhabha scattering at 183 and
189 GeV are computed with the program TOPAZ0~\cite{TOPAZ0} for 
total cross sections and forward-backward asymmetries in the
angular range of each experiment. This is sufficient for the
global analysis of the data from the DELPHI and L3 experiments.
We assign a theory uncertainty of 2~\% to the absolute scale of
the SM predictions. In order to analyze the differential cross
section measurements of the ALEPH and OPAL collaborations,
we compute the {\em form} of the differential spectra using
the generator BHWIDE~\cite{BHWIDE}, and then normalize the
total cross section to the TOPAZ0 prediction. Here a theory
error of 3~\% is assigned. In all cases the individual experimental
cuts of the selection procedures and the isolation of the high
energy samples are taken into account.

In total we have 44 data points: 36 from the 4 differential
spectra, 4 from the cross sections and 4 from the
forward-backward asymmetries. The effects coming from virtual
graviton exchange are computed as a function of the parameter
$\rm \varepsilon = \lambda/M_s^4$. A fitting procedure similar
to the one in~\cite{sneut} is applied.

\begin{figure}[htbp]
  \begin{center}
  \resizebox{0.85\textwidth}{0.60\textheight}{
  \includegraphics*{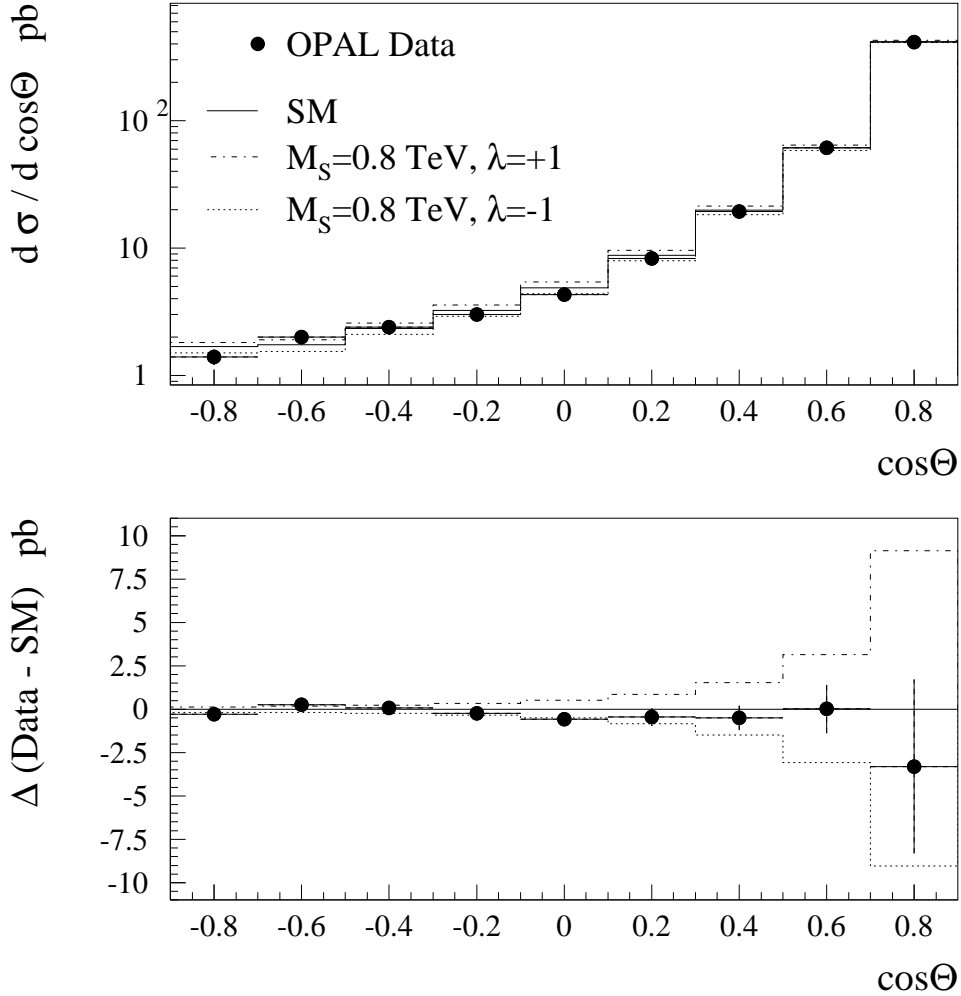}
  }
  \end{center}
  \caption{\em The differential cross section for Bhabha scattering
           at LEP2 in the Standard Model and models of low
           scale gravity. The data is from the OPAL collaboration
           at 189 GeV. The errors are statistical and systematic;
           the theory uncertainty is not shown.
           The lower plot shows the difference between 
           the data and the SM expectation together with the expected 
           deviations from the SM in models with large extra
           dimensions. The experimental sensitivity peaks in the
           forward direction.}
  \label{fig:figure1}
\end{figure}

A negative log-likelihood function is constructed by combining
all data points at the two centre-of-mass energies:
\Be
\rm -\log \mathcal{L} = \sum_{r=1}^{n}\left(\frac{(Prediction(SM, \varepsilon) - Measurement)^2}{2 \cdot \Delta_{Measurement}^2}\right)_r
\label{eqll1}
\Ee
\begin{eqnarray}
\rm \Delta_{Measurement}& = &\rm error(Prediction(SM, \varepsilon) - Measurement)\ , \nonumber \\
\label{eqll2}
\end{eqnarray}
where $ Prediction(SM, \varepsilon)$ is the SM expectation for the
given measurement (cross section or forward-backward asymmetry or a point
in the differential spectra) combined with 
the additional effect of graviton exchange as a function of the mass scale,
and $ Measurement$ is the corresponding measured quantity.
The index $\rm r$ runs over all data points.
The error on a deviation consists of three parts, which are combined in
quadrature: a statistical error and a systematic error (as given by the
experiments) and the theoretical error assigned above.

The result of the combined fit is:
\Be
\rm \varepsilon = -0.46\ ^{+0.37}_{-0.36} \ \ TeV^{-4}
\Ee
or 1.24 $\rm \sigma$ away from the Standard Model expectation
$\rm \varepsilon = 0$.

One example of the data analysis is shown in \Figref{figure1},
where the SM predictions and the expectations from the low
scale gravity model are compared to the measurements of the OPAL
collaboration at 189 GeV.

\begin{figure}[htbp]
  \begin{center}
  \resizebox{0.85\textwidth}{0.48\textheight}{
  \includegraphics*{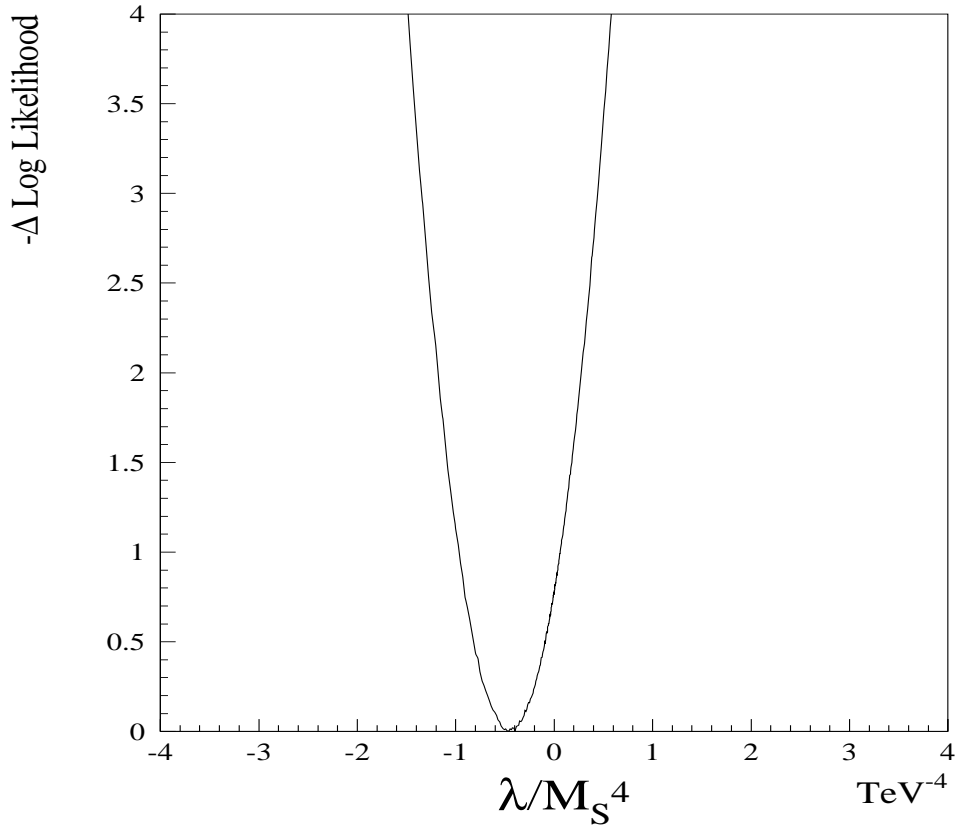}
  }
  \end{center}
  \caption{\em Log-likelihood curve of the global fit to the data on
           Bhabha scattering from the four LEP experiments.}
  \label{fig:figure2}
\end{figure}

So the data from the four LEP collaborations shows no statistically
significant deviations from the SM  predictions due to
the effects of virtual graviton exchange. In their absence, we use
the log-likelihood method to determine a one sided upper limit on the
scales $\rm M_s$ or $\rm \Lambda_T$ at the  95\% confidence level.
After proper normalization it gives the confidence level 
for any value of $\rm M_s$ in the physically allowed region.
The exact definition is
\Be
\rm \int_{0}^{\varepsilon^+_{95}}\mathcal{L}({\varepsilon '})d\varepsilon ' =
0.95\int_{0}^{\infty}\mathcal{L}({\varepsilon '})d\varepsilon '
\Ee
\Be
\rm \int_{-\varepsilon^-_{95}}^{0}\mathcal{L}({\varepsilon '})d\varepsilon ' =
0.95\int_{-\infty}^{0}\mathcal{L}({\varepsilon '})d\varepsilon '\ .
\Ee

The limits are summarized in Table~\ref{limit} and the combined
log-likelihood curve is shown in \Figref{figure2}.

The limits obtained here are higher than those in other global
fits to collider data~\cite{Gupta,Cheung}. In these papers the
best limits $\rm \sim$~1~TeV come from the TEVATRON data
on Drell-Yan production
due to the higher accessible centre-of-mass energies, but with
small data samples. In~\cite{Cheung} fermion-pair production at
LEP2 has been considered for all cases except Bhabha scattering.
The results presented here improve on the limits obtained by
individual LEP experiments~\cite{algr,l3gr}.

\begin{table}
 \renewcommand{\arraystretch}{1.25}
 \begin{center}
\begin{tabular}{|c|c|}
\hline
 $\lambda$ = -1 & $\lambda$ = +1 \\  
\hline
 \multicolumn{2}{|c|}{$\rm M_s$ [TeV]} \\
\hline
   0.962        &   1.261        \\
\hline
 \multicolumn{2}{|c|}{$\rm \Lambda_T$ [TeV]} \\
\hline
   1.077        &   1.412        \\
\hline
\end{tabular}
 \end{center}
 \caption{\em Limits on the gravity scales $\rm M_s$ and $\rm \Lambda_T$
          from Bhabha scattering at LEP2 at the 95~\% confidence level.}
  \label{limit}
\end{table}

At this point the alert reader may begin to worry: if Bhabha scattering
is such a sensitive tool to search for low scale gravity, maybe the
luminosity measurements of the LEP experiments, based on the very same
process, will also be affected and the results obtained here will not
be strictly valid. To close this loophole, a check is performed for a
typical angular range for luminosity measurements: for scattering
angles from 24 to 54 mrad. For a scale as low as $\rm M_s = 0.750\ TeV$,
excluded by this analysis, the change in the cross section is
0.008~\% for centre-of-mass energy of 188.7 GeV.

%
%
\section*{Conclusions}

The results of this work can be summarized as follows:
  \begin{itemize}
    \item Bhabha scattering is the golden channel to search for
          virtual graviton exchange at LEP2
    \item a global fit to the LEP2 data on Bhabha scattering is
          performed and limits on the mass scale of quantum
          gravity models with large extra dimensions of
          $\rm \Lambda_T = 1.077\ TeV$ for $\lambda = -1$ and
          $\rm \Lambda_T = 1.412\ TeV$ for $\lambda = +1$
          at 95~\% confidence level are set
    \item the precision of the Standard Model predictions for
          Bhabha scattering starts to be the limiting factor
          in this search; improved theory predictions are very desirable
          in view of the expected large data samples from the
          LEP running in 1999 and 2000.
  \end{itemize}

%
%
\section*{Acknowledgements}
The author is grateful to G.~Giudice, J.~Hewett, T.~Rizzo, S.~Mele,
E.~Sanchez and A.~Dominguez for discussions on some of the theoretical
and experimental aspects of this work.

%
%

\bibliographystyle{/l3/paper/biblio/l3stylem}
\bibliography{%
/l3/paper/biblio/l3pubs,%
/l3/paper/biblio/aleph,%
/l3/paper/biblio/delphi,%
/l3/paper/biblio/opal,%
/l3/paper/biblio/markii,%
/l3/paper/biblio/otherstuff,%
eth-pr-99-xx}

\end{document}